\documentclass[prd,showpacs,twocolumn]{revtex4}
\usepackage{epsfig}
\usepackage{amsmath}
\usepackage{amssymb}
\begin{document}
\title{  Annihilation Type Radiative Decays of $B$ Meson
in Perturbative QCD Approach}
\author{
 Ying Li\footnote{liying@mail.ihep.ac.cn}, Cai-Dian L\"u}
\affiliation{\it \small  CCAST (World Laboratory), P.O. Box 8730,
Beijing 100080, China;} \affiliation
 {\it \small Institute of High
Energy Physics, P.O.Box 918(4), Beijing 100049, China;}

\begin{abstract}
With  the  perturbative QCD approach based on $k_T$ factorization,
we study the   pure annihilation type  radiative decays $B^0 \to
\phi\gamma$ and $B^0\to J/\psi \gamma$. We find that the branching
ratio of $B^0 \to \phi\gamma$ is
$(2.7^{+0.3+1.2}_{-0.6-0.6})\times10^{-11}$, which is too small to
be measured in the current $B$ factories of BaBar and Belle. The
branching ratio of $B^0\to J/\psi \gamma$ is
$({4.5^{+0.6+0.7}_{-0.5-0.6}})\times10^{-7}$, which is just at the
corner of being observable in the $B$ factories. A larger
branching ratio $BR(B_s^0 \to J/\psi \gamma) \simeq 5 \times
10^{-6}$ is also predicted.  These decay modes will help us
testing the standard model and searching for new physics signals.
\end{abstract}
\maketitle
$B$ meson rare decays are interesting   for testing the standard
model and searching for new physics. However, due to our poor
knowledge of non-perturbative QCD, predictions for many
interesting exclusive decays are polluted by large hadronic
uncertainties.   The two body radiative $B$  decays involve
simpler hadronic dynamics with only one hadron in the final
states, so they suffer much less pollution
 than non-leptonic decays. The radiative decays such as
$B\to K^* \gamma$, $\rho (\omega) \gamma$  thus attract much
attention \cite{kpgamma}. The isospin breaking effects between the
charged $B^\pm$ and neutral $B^0$  in these modes are mainly due
to contributions from the annihilation type diagrams
\cite{kpgamma,hep-ph/0106067,hep-ph/0106081,hep-ph/0406055,
hep-ph/0508300}.

The importance of the annihilation type diagrams can also be shown
from the pure annihilation radiative $B$ decays. The color
suppressed $B^0\to J/\psi \gamma $ and $B^0 \to \phi\gamma$ modes
are of this kind. The former is tree dominant while the latter is a
pure penguin flavor changing neutral current decay. Despite the fact
that they are annihilation type decays, these decay amplitudes can
be factorized as the B meson to photon transition form factor
$\langle\gamma|\bar q \gamma_\mu (1-\gamma_5)b|B\rangle$ times the
decay constant $\langle\phi|\bar s \gamma_\mu s|0\rangle$ or
$\langle J/\psi|\bar c \gamma_\mu c|0\rangle$  in the naive
factorization approach.

Recently, the vertex corrections for the four quark operators have
been performed in the so called QCD factorization approach
\cite{bbns},  utilizing the light-cone wave functions. The
branching ratios turn out to be one order of magnitude different
from the naive factorization approach
\cite{hep-ph/0308256,hep-ph/0305283}. Such a large contribution
from next-to-leading order corrections implies that the hadronic
uncertainty in this kind of decays is as large as other hadronic
annihilation type decays \cite{annihilation}. Recent study of soft
collinear effective theory \cite{scet} also shows that the naive
factorization contribution is not the only dominant contribution,
which contradicts to QCD factorization claims. More theoretic
study is needed before one can claim new physics effects in these
decays.

In this paper, we will use an alternative approach- the perturbative
QCD approach (PQCD)  \cite{pqcd} to calculate the pure annihilation
type decays $B^0 \to \phi\gamma$ and $B^0\to J/\psi \gamma$. Based
on $\mathrm{k}_T$ factorization, the PQCD approach has been proposed
and applied to calculate two body non-leptonic $B$ decays such as $B
\to K\pi$ \cite{kpi}, $\pi\pi$ \cite{pipi}, $\rho\rho$
\cite{hep-ph/0508032},  the radiative decays $B \to K^*\gamma$
\cite{hep-ph/0406055}, $\rho (\omega)\gamma$ \cite{hep-ph/0508300},
 etc. and the results are consistent with experimental data. In
this approach, the quark transverse momentum $\mathrm{k}_T$ is
kept in order to kill  the end-point singularity. Because of
inclusion of transverse momenta, double logarithms  from the
overlap of two types of infrared divergences, soft and collinear,
are generated in radiative corrections. The resummation of these
double logarithms leads to a Sudakov form factor, which suppresses
the long-distance contribution.


For convenience, we work in the light-cone coordinate, where the $B$
meson   momentum in its rest frame, is
\begin{eqnarray}
P_B=(P_B^+,P_B^-,\vec{P}_{B\perp})=\frac{M_B}{\sqrt 2}( 1,1,\vec
0_\bot).
\end{eqnarray}
By choosing the coordinate frame where the vector meson moves in the
``$-$'' direction and photon in the ``+'' direction, the momenta of
final state particles are
\begin{eqnarray}
P_V &=&(P_V^+,P_V^-,\vec{P}_{V\perp}) =\frac{M_B}{\sqrt 2}(
r^2,1,\vec 0_\bot), \nonumber \\ P_{\gamma}
&=&(P_{\gamma}^+,P_{\gamma}^-,\vec{P}_{\gamma\perp})=
\frac{M_B}{\sqrt 2}( 1-r^2,0,\vec 0_\bot),
\end{eqnarray}
where $r=m_V/M_B$. The momentum of the light quark in $B$ meson
is:
\begin{eqnarray}
k_1&=&(k_1^{+},k_1^{-},\vec k_{1T})=(\frac{M_B}{\sqrt
2}x_1,0,\vec{k_{1T}}).
\end{eqnarray}
For the final state vector meson, we set the momentum of $q
(q=s,c)$ as
\begin{eqnarray}
k_2&=&(k_2^{+},k_2^{-},\vec k_{2T} )=(\frac{M_B}{\sqrt
2}x_2r^2,\frac{M_B}{\sqrt 2}x_2,\vec{k_{2T}}).
\end{eqnarray}
In above functions, $x_1$ and $x_2$ are momentum fractions of the
quarks.

\begin{figure}[thb]
\begin{center}
\includegraphics[scale=0.4]{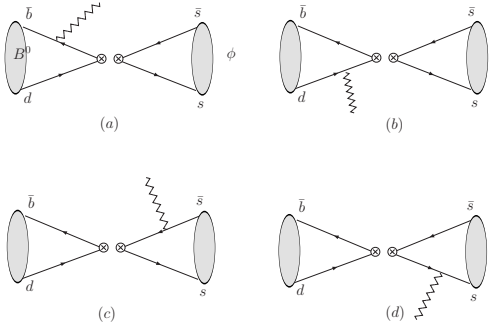}
\caption{Feynman diagrams for ${B}^0 \to \phi \gamma $ process in
PQCD.} \label{fig1}
\end{center}
\end{figure}

In PQCD approach, the decay amplitude is factorized into the
convolution of the mesons' wave functions, the hard scattering
kernel and the Wilson coefficients, which stand for the soft, hard
and harder dynamics respectively. With transverse momentum and
Sudakov form factor, the formalism can be written as:
\begin{multline}
\int^1_0 dx_1 dx_2 \int^{\infty}_0b_1 db_1
 b_2db_2\hspace{1mm}  \Phi_V(x_2,b_2)\Phi_B(x_1,b_1)
  \\
 \times C(t)
H(x_1,x_2,b_1,b_2,t) \exp{[-S(x_1,x_2,b_1,b_2,t)]},
\end{multline}
where  $b_i$ is the conjugate space coordinate of the transverse
momentum $k_{iT}$, which represents the transverse interval of the
meson. $t$ is chosen as the largest energy scale in the hard
scattering kernel $H$ in order to suppress higher order corrections.
The light cone wave functions of mesons are not calculable in
principle in PQCD, but they are universal for all the decay
channels. So that they can be constrained from the measured other
decay channels, like decays $B\to K\pi$ \cite{kpi}, $B\to \pi \pi$
\cite{pipi}, etc.

 Since the outgoing photon can be only transversely
polarized, the decay amplitude can be decomposed into two parts as:
\begin{equation}
A=(\varepsilon_{V}^*\cdot\varepsilon_{\gamma}^*)M^S+
\frac{i}{P_V\cdot P_{\gamma}}\epsilon_{\mu\nu\rho\sigma}
\varepsilon_{\gamma}^{*\mu}\varepsilon_{V}^{*\nu}
P_{\gamma}^{\rho}P_V^{\sigma}M^P, \label{MSP}
\end{equation}
where $P_V$ and $P_{\gamma}$ are the momenta of vector meson, and
photon, respectively.  $\varepsilon_{V}^*$ and
$\varepsilon_{\gamma}^*$ are the relevant polarization vectors.
The matrix element $M^{S(P)} $ can be calculated in the PQCD
approach.

In principal, the leading order contributions for  $B^0\to V\gamma$
($V=J/\psi, \phi$) decays involve only four-quark operators plus a
photon emitted from any of the quark line. The effective weak
Hamiltonian is formed by the $12$ four-quark operators and the
corresponding QCD corrected Wilson coefficients \cite{buras}. Very
recently  the electromagnetic penguin operator $O_{7\gamma}$
contribution through $B^0 \to \gamma\gamma$ with one photon
connecting to the $\phi$ meson is studied in ref.\cite{enh}. The
 branching ratio for $B\to \phi \gamma$ is found to be $1\times 10^{-11}$
 which is larger than
the four quark operator contribution from QCD factorization approach
\cite{hep-ph/0308256,hep-ph/0305283}. In PQCD language, the
contribution from $O_{7\gamma}$ is next-to-leading order. Its
contribution is still smaller than other contributions in PQCD
approach which will be shown later. The contribution of this kind of
operator   to the $B^0\to J/\psi \gamma$ decay is negligibly small.

The lowest order Feynman diagrams of $B^0\to \phi \gamma $ in PQCD
are shown in Fig.\ref{fig1}. In principle, the photon can be emitted
from any quark line of the four-quark operator. However, the
contribution of Fig.\ref{fig1}(c) is canceled exactly by that of (d)
because of topology symmetry. Thus, only the decay amplitudes from
Fig.\ref{fig1}(a) and (b) are left as:
\begin{multline} \label{AMPL}
M^{S}= \frac{1}{\sqrt{3}}G_F e M_B^3rf_V V_{\mathrm{CKM}} \int^1_0
dx_1\int^{\infty}_0 db_1b_1 \\ \times\phi_{B}(x_1,b_1)
 \Bigl[(1-r^2)C(t^a) K_0(b_1A_a)e^{-S_{B}(t^a)}\\
-(1-r^2)C(t^b) K_0(b_1B_a)e^{-S_{B}(t^b)} \Bigr],
\end{multline}
\begin{multline}\label{amp2}
M^{P}= \frac{1}{\sqrt{3}}G_F e M_B^3rf_VV_{\mathrm{CKM}} \int^1_0
dx_1\int^{\infty}_0 db_1b_1 \\
\times\phi_{B}(x_1,b_1)
 \Bigl[C(t^a) K_0(b_1A_a)e^{-S_{B}(t^a)}\\
+C(t^b) K_0(b_1B_a)e^{-S_{B}(t^b)} \Bigr],
\end{multline}
with
\begin{eqnarray}
A_a^2=(1+x_1-r^2)M_B^2, B_a^2=x_1(1-r^2)M_B^2, \nonumber
\end{eqnarray}
\begin{eqnarray}
t^a = \mbox{max}(A_a,1/b_1), ~~t^b = \mbox{max}(B_a,1/b_1).
\end{eqnarray}
$K_0(x)$ is the modified Bessel function which results from
Fourier transformation of the quark propagator. $e^{-S_B(t)}$ is
Sudakov form factor, $f_V$ is vector meson decay constant and
$V_{\mathrm{CKM}}$ denotes CKM matrix elements. In above
calculations, the $B$ meson is treated as a heavy-light system,
whose wave function is defined as:
\begin{eqnarray}
\Phi_{B} = \frac{i }{\sqrt{6}}(\not P_B +M_B)\gamma_5
 \phi_B(x_1,b_1),
\end{eqnarray}
and the expression of distribution amplitude $\phi_B$ is shown in
ref.\cite{kpi,pipi} with parameter $\omega_b=0.4$GeV, which is
normalized as
\begin{eqnarray}
\int_0^1 dx_{1}\phi_B (x_{1},b_1=0)&=&\frac{f_B}{2\sqrt{6}}\;,
\label{eq:bnor}
\end{eqnarray}
where   $f_B$ is the decay constant of the $B$ meson. Since we do
not need the $\phi$ or $J/\psi$ wave functions in the calculations
of Feynman diagrams Fig.\ref{fig1}(a) and (b), we need not show them
here.

For  $B^0\to \phi\gamma$ decay, only penguin operators can
contribute, and the CKM matrix elements are
$V_{\mathrm{CKM}}=V^*_{tb} V_{td}$. The
  combination of the Wilson coefficients in eq.(\ref{AMPL}, \ref{amp2})
are:
\begin{eqnarray}
C^P_{\phi\gamma}=C_3&+&\frac{1}{3}C_4+C_5+\frac{1}{3}C_6-\frac{1}{2}C_7
-\frac{1}{6}C_8 \nonumber \\
&-&\frac{1}{2}C_9-\frac{1}{6}C_{10}.
\end{eqnarray}
The Wilson coefficient of dominant QCD penguin operator $C_3$
cancels much with $C_4/3$, and $C_5$ cancels with $C_6/3$. This is a
result of color suppression in this decay, since at least three
gluons are needed for a vector $\phi$ meson produced from penguin
diagram \cite{xing}. For $B\to J/\psi\gamma$, both tree and penguin
operators give contribution, and corresponding conclusions of  the
Wilson coefficients are:
\begin{eqnarray}
C^T_{\psi\gamma}&=&C_1+\frac{1}{3}C_2, \nonumber \\
C^P_{\psi\gamma}&=&C_3+\frac{1}{3}C_4+C_5+\frac{1}{3}C_6+C_7\nonumber \\
&+&\frac{1}{3}C_8+C_9+\frac{1}{3}C_{10}.
\end{eqnarray}
Again, $C_1$ cancels much with $C_2/3$ which is also a result of
color suppressed tree contribution. The $V_{\mathrm{CKM}}$  in
 tree (penguin) operators is $V^*_{cb} V_{cd}$ ($V^*_{tb}
 V_{td}$).

 With the amplitudes $M^S$ and $M^P$ defined in Eq.(\ref{MSP}), the
  decay width of $B^0\to V\gamma$ is given by
\begin{equation}
\Gamma =\frac{|M^S|^2+|M^P|^2}{8\pi M_B} (1-r^2).
\end{equation}

 In our numerical calculations,  the input parameters
are summarized in Table~\ref{parameter}, where $\lambda$, $A$,
$\rho$ and $\eta$ are CKM parameters in Wolfenstein
parametrization \cite{wolfenstein}, and $\bar{\rho}=\rho
(1-\frac{1}{2}\lambda^2)$, $\bar{\eta}=\eta
(1-\frac{1}{2}\lambda^2)$. Their values can be found in
  Review of Particle Properties \cite{pdg}.

\begin{table}
\begin{center}
\caption{Summary of input parameters\cite{pdg}} \label{parameter}
\begin{tabular}{c}\hline\hline
CKM parameters and QCD constant\\
 $\begin{array}{cccccc}
 \lambda & A &\bar{\rho}& \bar{\eta}&\Lambda_{\overline{\mathrm{MS}}}^{(f=4)}&\tau_{B^0}\\
 0.2196& 0.819&0.20& 0.33& 250 \mbox{MeV}&1.54\mbox{ps}
 \end{array}$\\
 \hline
Meson decay constants \\
 $\begin{array}{cccc}
   f_B &f_{B_s} & f_{\phi} &f_{J/\psi} \\
  216\mbox{MeV}&236\mbox{MeV}& 254\mbox{MeV} & 405 \pm 14\mbox{MeV}
 \end{array}$ \\\hline
  Meson masses\\
 $\begin{array}{cccc}
    M_W&  M_B  & M_{\phi} & M_{J/\psi}\\
  80.41\mbox{GeV} &  5.28\mbox{GeV}&1.02\mbox{GeV}&3.10\mbox{GeV}
 \end{array}$ \\
   \hline\hline
\end{tabular}
\end{center}
\end{table}

 At the leading order, the main
uncertainty for decay branching ratios comes from the $B$ meson
wave function. But it is constrained by the measured exclusive
hadronic decays, like $B \to K \pi $ \cite{kpi}, $B \to\pi\pi$
\cite{pipi} with parameter $\omega_B$ from $0.3 \mathrm{GeV}$ to
$0.5 \mathrm{GeV}$. On the other hand, there should be large
uncertainty since we work only at leading order in $\alpha_s$ for
the hard part and also for the Wilson coefficients. The missing
next-to-leading order correction is a very important uncertainty
for rare decays. To estimate it, we consider the hard scale $t$ at
a range of
\begin{eqnarray}
&&\max(0.75A_a, \frac{1}{b_1})<t^a<\max(1.25A_a,
 \frac{1}{b_1}),\\
&&\max(0.75B_b,\frac{1}{b_1})<t^b<\max(1.25B_b,
 \frac{1}{b_1}).
\end{eqnarray}
With the above major uncertainties from $B$ meson wave function
parameter $\omega_B$ and the different scale $t$, respectively, we
 give out the branching ratio of $B^0\to \phi\gamma$:
\begin{equation}
\mathrm{BR}(B^0\to
\phi\gamma)=(2.7^{+0.3+1.2}_{-0.6-0.6})\times10^{-11}.\label{br}
\end{equation}
There are many other uncertainties in our calculation such as
decay constants and CKM matrix elements. However, the uncertainty
induced by above factors is not more than
$10\%$\cite{hep-ph/0605112}. With such a small branching ratio
(\ref{br}), this decay is too rare to be measured at the running
$B$ factories or even in future LHC-b experiment. Our result is
still much smaller than the recent upper limit $8.5\times10^{-7}$
from experiments \cite{hep-ex/0501038}. If some new physics
particles enhance this ratio through tree or loop effects
\cite{hep-ph/0308256}, it  may be measurable in near future
experiments.

After similar calculations, we also give the branching ratio of
$B^0\to J/\psi\gamma$ decay:
\begin{equation}
\mathrm{BR}(B^0\to
J/\psi\gamma)=({4.5^{+0.6+0.7}_{-0.5-0.6}})\times10^{-7}.
\end{equation}
Because this process is tree diagram dominated, it is much larger
than that of $B^0\to \phi\gamma$. This result is still not big
enough to be measured at the $B$ factories but it is already
around the corner of  $B$ factories capability. Currently, there
is only upper limit, which is $1.6 \times 10^{-6}$ at $90\%$
confidence level \cite{hep-ex/0408018}.

From our calculation, we find that most of the contribution comes
from Fig.1(b), which indicates that photon emission from the light
quark of the $B$ meson is easier than that from the heavy quark.
The reason is that the heavy quark is much difficult to become
off-shell than the light quark, while a nearly on-shell quark
rarely emits photons.

Within QCD factorization approach, the branching ratios of these two
decays have been calculated in
Ref.\cite{hep-ph/0308256,hep-ph/0305283}. The results are given as :
\begin{eqnarray}
\mathrm{BR}(B^0\to J/\psi\gamma)=7.65\times10^{-9}
;\\
\mathrm{BR}(B^0\to \phi\gamma)=3.6\times10^{-12}.
\end{eqnarray}
Compared with their results, we find that our results are one
order of magnitude larger. In QCD factorization approach, some
next-to-leading order contribution has been added in addition to
the naive factorization contribution. It shows that these
presumably order $\alpha_s$ corrections change the branching
ratios by one order of magnitude. In fact these two decays are
non-factorizable contribution dominant in naive factorization
approach. Usually QCD factorization approach calculation for this
kind of decays (such as $B^0 \to \bar D^0 \pi^0$ \cite{dpi}) is
not stable.
 They receive dominant contribution from non-factorizable diagrams,
sometimes with an endpoint singularity in the collinear
factorization. In fact, our numerical results are also much larger
than the naive factorization approach, which is mainly due to a
smaller energy scale ($\sqrt{\Lambda_{QCD}m_b}$ other than $m_b$)
used in the calculation of the Wilson coefficients. This
uncertainty comes from part of the next-to-leading order effect in
PQCD approach.

The $B_s \to J/\psi \gamma$ decay is the same as $B^0 \to J/\psi
\gamma$ decay within SU(3) symmetry. It is also a pure annihilation
type decay. The decay formulas are exactly the same except
   replacing the corresponding CKM factor $V_{cd}$, $V_{td}$
 by $V_{cs}$, $V_{ts}$, respectively. Taking into account the SU(3) breaking effect,
 we used $\omega_{B_s}
= 0.5 \mathrm{GeV}$ \cite{bs}, $f_{B_s}=236$MeV, we can get
\begin{eqnarray}
BR(B_s \to J/\psi \gamma)\simeq 5 \times 10^{-6},
\end{eqnarray}
 which should be easy to be measured in the
future LHCb experiment.


  In a summary, we calculate the branching ratios of pure
annihilation type radiative decays $B^0\to \phi\gamma$ and $B^0\to
J/\psi\gamma$ within the standard model in PQCD approach. We find
the branching ratio of $B^0\to \phi\gamma$ is at the order of
$10^{-11}$. This small branching can not be detected in the running
B factories of BaBar and KEK, unless some new physics enhance this
results sharply. In future, this decay may be measured in LHC-b
experiment or other high luminosity experiments. For $B^0 \to J/\psi
\gamma$ decay, the branching ratio is about $4\times 10^{-7}$, which
is just close to the $B$ factory experiment capability of
measurement.
 The experimental measurements of these decays would be very useful for
understanding various QCD methods, like QCD factorization and PQCD
approaches.

\section*{Acknowledgments}
This work was partly supported by the National Science Foundation of
China. We thank Y.D. Yang, J.-X Chen, Y.-L Shen, W. Wang, X.-Q Yu
and J. Zhu for useful discussions.

\end{document}